\documentclass{ws-procs975x65}

\begin{document}

\title{Some new perspectives on pairing in nuclei}

\author{S.~Pittel}

\address{Bartol Research Institute, University of
Delaware, Newark, Delaware 19716, USA\\
}

\author{J. Dukelsky}

\address{Instituto de Estructura de la Materia, Consejo Superior de
Investigaciones Cientificas, Serrano 123, 28006 Madrid, Spain\\
}


\maketitle

\abstracts{ Following a brief reminder of how the pairing model
can be solved exactly, we describe how this can be used to address
two interesting issues in nuclear structure physics. One concerns
the mechanism for realizing superconductivity in finite nuclei and
the other concerns the role of the nucleon Pauli principle in
producing $sd$ dominance in interacting boson models of nuclei.}

\section{Introduction}

Ever since the work of Richardson in the mid-60s\cite{Richardson},
it has been recognized that the Pairing Model (PM) is exactly
solvable, even in the presence of non-degenerate single-particle
levels.  In recent years, there has been a resurgence of interest
in the PM, with several applications reported that build on this
exact solvability \cite{sie1}.

This talk reviews two recent applications of the PM in nuclear
physics. Both build on the fact that there exists a classical
electrostatic analogy for every PM. One makes use of this analogy
to obtain a pictorial representation of how superconductivity
arises in finite nuclear systems \cite{2DCoulomb}. The other has
led us to propose a new mechanism for $sd$ dominance in
interacting boson models of nuclei\cite{IBM}.

\section{Richardson's solution of the Pairing Model}

The PM hamiltonian for both fermions and bosons can be written as

\begin{equation}
H_P~=~ \sum_l \epsilon_l \hat{N}_l ~+~ \frac{g}{2} \sum_{ll'}
A^{\dagger}_l A_l ~, \label{PM}
\end{equation}
where
\begin{equation}
\hat{N}_l ~=~ \sum_m a^{\dagger}_{lm}a_{lm} ~~~,~~~ A^{\dagger}_l
~=~ \sum_m a^{\dagger}_{lm} a^{\dagger}_{l\bar{m}} ~. \label{NA}
\end{equation}
Here $a^{\dagger}_{lm}$ creates either a boson or a fermion in
single-particle state $lm$ and ${l\bar{m}}$ denotes the time
reverse of $lm$.

Richardson considered the following ansatz for the ground state of
a system of $2N$ particles subject to this hamiltonian:

\begin{equation}
|\Psi \rangle ~=~ \prod_{i=1}^N B^{\dagger}_{\alpha} ~ |0\rangle
~~, ~~ B^{\dagger}_{\alpha}~=~ \sum_l \frac{1}{2\epsilon_l
-e_{\alpha}} ~ A^{\dagger}_l ~. \label{ansatz1}
\end{equation}
He showed that it is an exact eigenstate of the pairing
hamiltonian if the {\it pair energies} ${\it e_{\alpha}}$ satisfy
the set of equations ($\Omega_l = l +\frac{1}{2}$)

\begin{equation}
1 ~+~ 2g \sum_l \frac{\Omega_l}{2\epsilon_l -e_{\alpha}} ~\mp ~ 4g
\sum_{\beta(\neq \alpha)} \frac{1}{e_{\beta}-e_{\alpha}} ~=~ 0 ~.
\label{Rich}
\end{equation}
In eq. (\ref{Rich})  and throughout the presentation, the upper
sign refers to boson systems and the lower sign to fermion
systems.

The coupled equations (\ref{Rich}), one for each of the $N$
collective pairs, are called the Richardson equations.

Once the set of Richardson equations has been solved, the total
ground state energy of the system can be obtained by summing the
resulting pair energies,

\begin{equation}
E~=~ \sum_{\alpha} e_{\alpha} ~.
\end{equation}

While the above discussion focused on the ground state solution,
it is possible to use the same general procedure to generate {\it
all} excited states as well.

\section{An electrostatic analogy for Pairing Models}

As we have seen, the eigenvalues and eigenfunctions of the
hamiltonian can be obtained using the Richardson approach, both
for fermion and boson systems. From this, it is straightforward to
establish an exact electrostatic analogy for the quantum pairing
problem. To do so, consider the energy functional

\begin{eqnarray}
U &=& \mp~\frac{1}{4g} \left[\sum_{\alpha }e_{\alpha } - \sum_{j}
\Omega_j \epsilon _{j} \right]  +\frac{1}{2}\sum_{j\alpha }(\pm
\Omega_{j})\ln \left| 2\epsilon
_{j}-e_{\alpha}\right|   \nonumber \\
&-&\frac{1}{2}\sum_{\alpha \neq \beta }\ln \left| e_{\alpha
}-e_{\beta }\right| -\frac{1}{8}\sum_{i\neq j} \Omega_{i}
\Omega_{j}\ln \left| 2\epsilon _{i}-2\epsilon _{j}\right|.
\end{eqnarray}

It can be readily shown that when we differentiate $U$ with
respect to the pair energies $e_{\alpha}$ and equate  to zero we
recover precisely the Richardson equations (\ref{Rich}).

To appreciate the physical meaning of $U$, we should remember that
the Coulomb interaction between two point charges in two
dimensions is
\begin{equation}
v\left( {\bf r}_{1},{\bf r}_{2}\right) =-q_{1}q_{2}\ln \left| {\bf r}_{1}-%
{\bf r}_{2}\right| ~,
\end{equation}
where $q_{i}$ is the charge and $r_{i}$ the position of particle
$i$.

Thus, $U$ is the energy functional for a classical two-dimensional
(2D) electrostatic system with the following ingredients:

\begin{itemize}
\item
There are a set of {\em fixed charges}, one for each
single-particle level, which are located at the positions
$2\epsilon _{i}$ and have charges $\pm\frac{\Omega_i}{2}$. We will
call them {\em orbitons}.

\item
There are $N$ {\em free charges}, one for each collective pair,
which are located at the positions $e_{\alpha }$ and have positive
unit charge. We will call them {\em pairons}.
\item
There is a Coulomb interaction between all charges.
\item
There is a uniform electric field in the vertical direction with
intensity $\pm\frac{1}{4g}$.
\end{itemize}

The existence of this exact analogy suggests that we might be able
to use the positions that emerge for the pairons in the classical
problem to gain insight into the quantum problem, hopefully
insight that was not otherwise evident.

Some other properties of the electrostatic problem that we will be
using are:

\begin{itemize}

\item
Since the orbiton positions are given by the single-particle
energies, which are real, they must lie on the vertical or real
axis.

\item
For fermion problems, the pair energies that emerge from the
Richardson equations are not necessarily real. They can either be
real or they can come in complex conjugate pairs. Thus, a pairon
must either lie on the vertical axis (real pair energies) or be
part of a mirror pair (complex pair energies).
\item
For boson problems, the pair energies are of necessity real and,
thus, like the orbitons lie on the real axis.

\end{itemize}

\section{A new pictorial representation of nuclear superconductivity}

We now apply the electrostatic analogy to the problem of identical
nucleon pairing and in particular to the question of how
superconductivity arises in such systems. Because of the limited
number of active nucleons in a nucleus, it is extremely difficult
to see evidence for the transition to superconductivity in such
systems.

We will discuss what happens when we apply the electrostatic
analogy to the semi-magic nuclei $^{114-116}Sn$. The calculations
are done as a function of pairing strength $g$, using
single-particle energies extracted from experiment. Table 1 shows
the corresponding information on the positions and charges of the
orbitons.

\begin{table}[t]
\caption{Position and charges of the orbitons appropriate to a
pairing treatment of $^{114-116}Sn$.\label{tab:Sn}} \vspace{0.1cm}
\begin{center}
\footnotesize
\begin{tabular}{|c|c|c|}
\hline
  Orbiton & Position & Charge \\
  \hline
\hline
$d_{5/2}$ & $0.0$ & $-1.5$ \\
 \hline
 $g_{7/2}$ & $0.44$ & $-2.0$
\\
\hline
$s_{1/2}$ & $3.80$ & $-0.5$ \\
\hline
  $d_{3/2}$ & $4.40$ & $-1.0$
\\
\hline
$h_{11/2}$ & $5.60$ & $-3.0$ \\
\hline
\end{tabular}
\end{center}
\end{table}

Fig. 1 focuses on the nucleus $^{114}Sn$, showing the positions of
the pairons in the 2D plane as a function of $g$. Since $^{114}Sn$
has 14 valence neutrons, there are seven pairons in the classical
picture. In the limit of very weak coupling, six neutrons fill the
$d_{5/2}$ orbit and eight fill the $g_{7/2}$. The corresponding
electrostatic picture (Fig. 1a) has three pairons close to the
$d_{5/2}$ orbiton and four close to the $g_{7/2}$.  In the figure,
we draw lines connecting each pairon to the one that is closest to
it. These lines make clear that at very weak coupling the pairons
organize themselves as artificial {\em atoms} around their
corresponding orbitons.

\begin{figure}
\centerline{\epsfxsize=4in\epsfbox{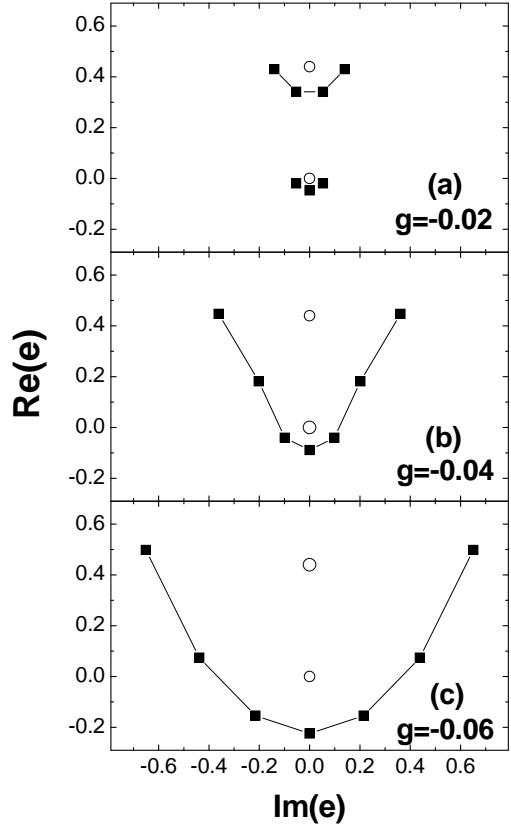}} \vspace{-1cm}
\caption{Two-dimensional representation of the pairon positions in
$^{114}Sn$ for three selected values of $g$. The orbitons are
represented by open circles; only the lowest two, the $d_{5/2}$
and $g_{7/2}$, are shown at the positions dictated by Table 1.
\label{fig1}}
\end{figure}

What happens as we increase the magnitude of $g$ (Figs. 1b-c)?
[The physical value is roughly $-0.092~MeV$.] As $g$ increases,
the pairons repel, causing the atoms to expand. For $g \approx
-0.04$, a transition takes place from two isolated atoms to a {\em
cluster}, with all pairons connected to one another. We claim that
this geometrical transition from atoms to clusters in the
classical problem is a reflection of the superconducting
transition in the quantum problem.

We have also treated the nucleus $^{116}Sn$, with the same set of
single-particle energies as in $^{114}Sn$. What we find is that in
$Sn^{116}$ the transition to complete superconductivity occurs in
two stages. For small $g$, the pairons distribute themselves into
three atoms, surrounding the $d_{5/2}$, $g_{7/2}$ and $s_{1/2}$
orbitons. When $g$ reaches  roughly $-0.06$, the two lowest atoms
- containing 7 pairons - merge into a cluster, as in $^{114}Sn$,
with the eighth still separate. When $g$ grows to roughly $-0.095$
a second transition takes place, with the eighth pairon merging
into a larger cluster with the other seven. From this point on,
superconductivity is complete.

\section{A new mechanism for $sd$ dominance in the IBM}

The electrostatic analogy can also be applied to boson pairing
models, with the important caveat that the pairons are now
confined to the real axis. Fig. 2 shows the pairon positions for a
model involving $10$ bosons moving in all even-$L$ boson states up
to $L=12$ and interacting via repulsive boson pairing with
strength $g$. The single-boson energies are assumed to increase
linearly with $l$.

Several points are immediately apparent. At low pairing strength,
the pairons sit very near the $s$ orbiton, reflecting the fact
that the bosons are almost completely in the $s$ state. As the
pairing strength increases, a phase transition takes place to a
scenario in which the pairons are no longer sitting near the $s$
orbiton. However, even after the phase transition all pairons are
confined to the region between the lowest two orbitons, the $s$
and $d$. What this suggests is that after the phase transition the
boson pairs that define the corresponding quantum ground state are
most likely primarily of $s$ and $d$ character.

%%%%%%%%%%%%%%%%%%%%%%%%%%%%%%%%%%%%%%%%%%%%%%%%
\begin{figure}

\centerline{\epsfxsize=4in\epsfbox{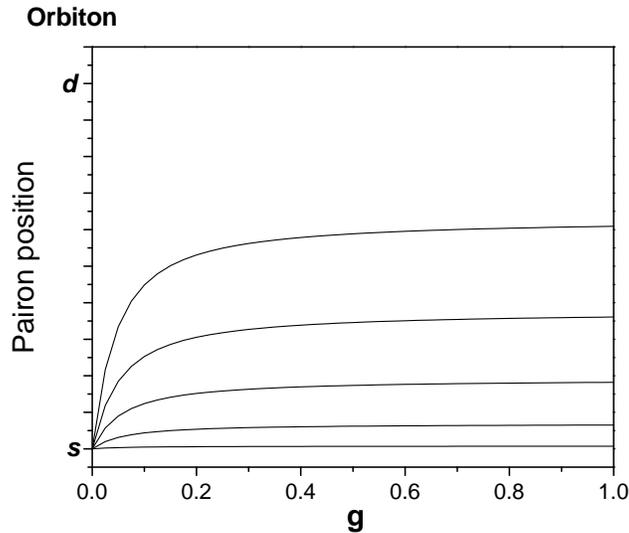}}
\vspace{-2cm}\caption{Evolution of pairons for a model involving
10 bosons in all even-$L$ states up to $L=12$ subject to a
hamiltonian with linear single-boson energies and a repulsive
boson pairing interaction \label{fig2}}
\end{figure}

What is the relevance of this to the IBM? As a reminder, in the
IBM the $s$ and $d$ bosons model the lowest two pair degrees of
freedom for identical nucleons, those with $J^{\pi}=2^+$ and
$4^+$. The key assumption of the model is that all other bosons,
reflecting higher pair states, can be ignored, except for their
renormalization effects. A second point to remember is that in any
effort to model composite objects by structureless particles,
there invariably arises a repulsive interaction between these
particles, to reflect the Pauli exchange between their
constituents.

The results in Fig. 2 are suggestive that in the presence of such
a repulsive interaction between bosons only the two lowest boson
degrees of freedom can correlate, namely the $s$ and $d$. This
suggests that repulsive pairing between bosons provides a new
mechanism for $sd$ dominance in interacting boson models of
nuclei.

%%%%%%%%%%%%%%%%%%%%%%%%%%%%%%%%%%%%%%%%%%%%%%%%
\begin{figure}
\centerline{\epsfxsize=4in\epsfbox{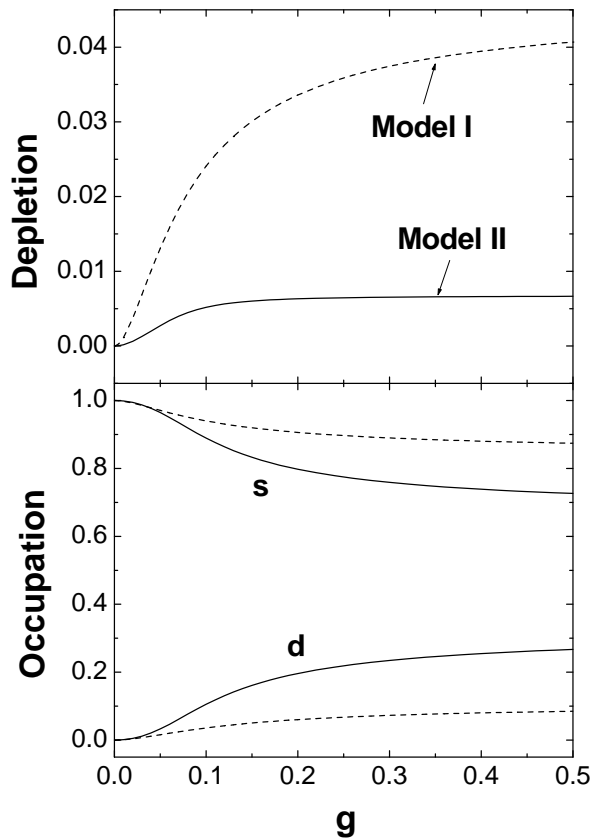}}
\vspace{-1cm}\caption{Occupation probabilities for the ground
state of a system of $5$ boson pairs and maximum angular momentum
$L=12$ as a function of the $g$. The upper graph shows the sum of
occupation probabilities (depletion) for high-spin bosons ($l>2$)
while the lower graph gives the occupation probabilities for $s$
and $d$ bosons. The dashed lines refer to Model I and the solid
line4s to Model II, as described in the text. \label{fig3}}
\end{figure}

These points can be made more quantitative by looking directly at
the quantum results. In Fig. 3, we show results for the same
interacting boson model as above, but now with two possible
choices for the single-boson spectrum. In addition to the choice
$\epsilon_l = l$ used before (Model I), we also consider
$\epsilon_l=l^2$ (Model II). In this way, we can assess whether
$sd$ dominance is a general feature of boson models involving
repulsive pairing or is limited to the model earlier shown.

As we can see from the figure, both models show the same general
features. For weak $g$, most of the bosons are in the $s$ state.
As $g$ increases, there is a phase transition to a mixed or
fragmented state. However, even in the fragmented state there are
essentially no bosons other than those with $L=0$ and $2$.

Indeed, when we carry out the calculation as a function of boson
number, we find that as $N$ grows the number of non-$sd$ bosons
decreases, and in the thermodynamic limit there are only $s$ and
$d$ bosons.

\section{Summary}

There are two key points we have tried to get across in this
presentation. The first is that pairing models, even with
non-degenerate levels, can be solved exactly using a method
introduced by Richardson almost 40 years ago. The second is that
these exactly solvable models can be used to provide interesting
insight into several issues of importance in nuclear physics. The
two examples we discussed concerned the mechanism for realizing
superconductivity in finite nuclear systems and the role of the
nucleon Pauli principle in producing $sd$ dominance in interacting
boson models of nuclei.

\section*{Acknowledgments}
This work was supported in part by the US National Science
Foundation under grant \#s PHY-9970749 and PHY-0140036 and by the
Spanish DGI under grant BFM2000-1320-C02-02.

\end{document}